# An Approach Finding Frequent Items In Text Or Transactional Data Base By Using BST To Improve The Efficiency Of Apriori Algorithm

P.Vasanth Sena
Assoc Prof in IT-Dept
Sree Chaitanya College Of Engineering
Karimnagar, India
E-mail: vasanthmtechsit521@gmail.com

T.Bhaskar
Sr.Asst Prof in IT-Dept
Sree Chaitanya College Of Engineering
Karimnagar, India
E-mail: bhas_ani@yahoo.co.in

K.Rmakrishna
Assoc Prof in CSE-Dept
Aizza College Of Engg & Technology
Mancherial, India

*Abstract:* Data mining techniques have been widely used in various applications. Binary search tree based frequent items is an effective method for automatically recognize the most frequent items, least frequent items and average frequent items. This paper presents a new approach in order to find out frequent items. The word frequent item refers to how many times the item appeared in the given input. This approach is used to find out item sets in any order using familiar approach binary search tree. The method adapted here is in order to find out frequent items by comparing and incrementing the counter variable in existing transactional data base or text data. We are also representing different approaches in frequent item sets and also propose an algorithmic approach for the problem solving.

## I.INTRODUCTION

A variety of data mining problems have been studied to help people to get insight information in the huge amount of data. We explored some of the algorithms in order to find out frequent item sets such as the appriori algorithm, in order to improve the efficiency of apriori frequent pattern tree (FP-tree), Mining frequent item sets using vertical data format. In this paper we mainly concentrate how to construct binary search tree to find out frequent item sets for the given transaction data base or text data base. A set of items is referred to as item set. An item set that contains k items is called k-item set. For example {bike, helm ate} is two item set. The frequency occurrence of items is number of time appeared in transactional data base, occurrence of word in text data base.

We need to take all the item sets from the transactional data base or all the text data, find out frequent items we construct binary search tree according to the rules of i.e the value of the left child always less than the root, value at the right child greater than the root, if there is any sub tree it must be binary search tree.

**Searching frequent 1-itemsets:** The first iteration of Apriori, whose pseudo code is shown in below, is very simple. F1 is optimally built by counting all the occurrence of each item i ϵ {1,2,…m} in every t ϵ D. an Array of m positions is used to store the item counters. Occurrences are counted by scanning D, and the items having minimum support are included into F1.

1: for all i | 1≤i≤m do
2: COUNT[i] ← 0
3: end for
4: for all t ϵ D do
5: for all i ϵ t do
6: COUNT[i] ← COUNT[i]+1
7: end for
8: end for
9: F1={ i | 1≤i≤m | CONT[I] ≥MIN_SUP}

## II.LITERATURE SURVEY:

Apriori is a seminal algorithm proposed by R. Agrawal and R. Srikant in 1994 for mining frequent item sets for Boolean association rules. The name of the algorithm is based on the fact that the algorithm uses *prior knowledge* of frequent item set properties, as we shall see following. Apriori employs an iterative approach known as a *level-wise* search, where $k$-item sets are used to explore $(k+1)$-item sets. First, the set of frequent 1-item sets is found by scanning the database to accumulate the count for each item, and collecting those items that satisfy minimum support. The resulting set is denoted $L1$. Next, $L1$ is used to find $L2$, the set of frequent 2-itemsets, which is used to find $L3$, and so on, until no more frequent $k$-itemsets can be found. The finding of each $Lk$ requires one full scan of the database. To improve the efficiency of the level-wise generation of frequent itemsets, an important property called the Apriori property, presented below, is used to reduce the search space. We will first describe this property, and then show an example illustrating its use. Apriori property: *All nonempty subsets of a frequent itemset must also be frequent.* The Apriori property is based on the following observation. By definition, if an itemset $I$ does not satisfy the minimum support threshold, *min sup*, then $I$ is not frequent; that is, $P(I) < min\ sup$. If an item $A$ is added to the itemset $I$, then the resulting itemset (i.e., $I[A]$ cannot occur more frequently than $I$. Therefore, $I[A]$ is not frequent either; that is, $P(I[A]) < min\ sup$. This property belongs to a special category of properties called antimonotone in the sense that *if a set cannot pass a test, all of its supersets will fail the same test as well*. It is called *antimonotone* because the property is monotonic in the context of failing a test.7 *"How is the Apriori property used in the algorithm?"* To understand this, let us look at how $Lk-1$ is used to find $Lk$ for $k \geq 2$. A two-step process is followed, consisting of join and prune actions.

1. The join step: To find $Lk$, a set of candidate $k$-itemsets is generated by joining $Lk-1$ with itself. This set of candidates is denoted $Ck$. Let $l1$ and $l2$ be itemsets in $Lk-1$. The notation $li[j]$ refers to the $j$th item in $li$ (e.g., $l1[k-2]$ refers to the second to the last item in $l1$). By convention, Apriori assumes that items within a transaction or itemset are sorted in lexicographic order. For the $(k-1)$-itemset, $li$, this means that the items are sorted such that $li[1] < li[2] < \ldots < li[k-1]$. The join, $Lk-1$ on $Lk-1$, is performed, where members of $Lk-1$ are joinable if their first $(k-2)$ items are in common. That is, members $l1$ and $l2$ of $Lk-1$ are joined if $(l1[1] = l2[1]) \wedge (l1[2] = l2[2]) \wedge \ldots \wedge (l1[k-2] = l2[k-2]) \wedge (l1[k-1] < l2[k-1])$. The condition $l1[k-1] < l2[k-1]$ simply ensures that no duplicates are generated. The resulting itemset formed by joining $l1$ and $l2$ is $l1[1], l1[2], \ldots, l1[k-2], l1[k-1], l2[k-1]$.

2. The prune step: $Ck$ is a superset of $Lk$, that is, its members may or may not be frequent, but all of the frequent $k$-itemsets are included in $Ck$. A scan of the database to determine the count of each candidate in $Ck$ would result in the determination of $Lk$ (i.e., all candidates having a count no less than the minimum support count are frequent by definition, and therefore belong to $Lk$). $Ck$, however, can be huge, and so this could

Transactional data for an *All Electronics* branch.

| TID | List of item IDs |
|---|---|
| T100 | I1, I2, I5 |
| T200 | I2, I4 |
| T300 | I2, I3 |
| T400 | I1, I2, I4 |
| T500 | I1, I3 |
| T600 | I2, I3 |
| T700 | I1, I3 |
| T800 | I1, I2, I3, I5 |
| T900 | I1, I2, I3 |

*FIG:1*

Involve heavy computation. To reduce the size of $Ck$, the Apriori roperty is used as follows. Any $(k-1)$-

itemset that is not frequent cannot be a subset of a frequent $k$-itemset. Hence, if any $(k-1)$-subset of a candidate $k$-itemset is not in $L_{k-1}$, then the candidate cannot be frequent either and so can be removed from $C_k$. This subset testing can be done quickly by maintaining a hash tree of all frequent itemsets.

Example : Apriori. Let's look at a concrete example, based on the *AllElectronics* transaction database, $D$, of Table 5.1. There are nine transactions in this database, that is, $|D| = 9$. We use Figure 5.2 to illustrate the Apriori algorithm for finding frequent itemsets in $D$.

In the first iteration of the algorithm, each item is a member of the set of candidate 1-itemsets, $C_1$. The algorithm simply scans all of the transactions in order to count the number of occurrences of each item.

1. Suppose that the minimum support count required is 2, that is, *min sup* = 2. (Here, we are referring to *absolute* support because we are using a support count. The corresponding relative support is 2/9 = 22%). The set of frequent 1-itemsets, $L_1$, can then be determined. It consists of the candidate 1-itemsets satisfying minimum support. In our example, all of the candidates in $C_1$ satisfy minimum support.

2. To discover the set of frequent 2-itemsets, $L_2$, the algorithm uses the join $L_1$ on $L_1$ to generate a candidate set of 2-itemsets, $C_2$.8 $C_2$ consists of $|L_1|$ 2_2-itemsets. Note that no candidates are removed from $C_2$ during the prune step because each subset of the candidates is also frequent

3. Next, the transactions in $D$ are scanned and the support count of each candidate item set in $C_2$ is accumulated, as shown in the middle table of the second row in Figure 5.2.

4. The set of frequent 2-itemsets, $L_2$, is then determined, consisting of those candidate 2-itemsets in $C_2$ having minimum support.

5. The generation of the set of candidate 3-itemsets, $C_3$, is detailed in Figure 5.3. From the join step, we first get $C_3 = L_2$ on $L_2$ = ff$I_1$, $I_2$, $I_3$g, f$I_1$, $I_2$, $I_5$g, f$I_1$, $I_3$, $I_5$g, f$I_2$, $I_3$, $I_4$g, f$I_2$, $I_3$, $I_5$g, f$I_2$, $I_4$, $I_5$gg. Based on the Apriori property that all subsets of a frequent item set must also be frequent, we can determine that the four latter candidates cannot possibly be frequent. We therefore remove them from $C_3$, thereby saving the effort of unnecessarily obtaining their counts during the subsequent scan of $D$ to determine $L_3$.

6. Note that when given a candidate $k$-itemset, we only need to check if its $(k-1)$-subsets are frequent since the Apriori algorithm uses a level-wise search strategy. The resulting pruned version of $C_3$ is shown in the first table of the bottom row of Figure 5.2.

7. The transactions in $D$ are scanned in order to determine $L_3$, consisting of those candidate 3-itemsets in $C_3$ having minimum support (Figure: 1)
    a. Join: $C_3 = L_2$ on $L_2$ = ff$I_1$, $I_2$g, f$I_1$, $I_3$g, f$I_1$, $I_5$g, f$I_2$, $I_3$g, f$I_2$, $I_4$g, f$I_2$, $I_5$gg on ff$I_1$, $I_2$g, f$I_1$, $I_3$g, f$I_1$, $I_5$g, f$I_2$, $I_3$g, f$I_2$, $I_4$g, f$I_2$, $I_5$gg = ff$I_1$, $I_2$, $I_3$g, f$I_1$, $I_2$, $I_5$g, f$I_1$, $I_3$, $I_5$g, f$I_2$, $I_3$, $I_4$g, f$I_2$, $I_3$, $I_5$g, f$I_2$, $I_4$, $I_5$gg.
    b. Prune using the Apriori property: All nonempty subsets of a frequent itemset must also be frequent. Do any of the candidates have a subset that is not frequent?
    c. The 2-item subsets of f$I_1$, $I_2$, $I_3$g are f$I_1$, $I_2$g, f$I_1$, $I_3$g, and f$I_2$, $I_3$g. All 2-item subsets of f$I_1$, $I_2$, $I_3$g are members of $L_2$. Therefore, keep f$I_1$, $I_2$, $I_3$g in $C_3$.

d. The 2-item subsets of fI1, I2, I5g are fI1, I2g, fI1, I5g, and fI2, I5g. All 2-item subsets of fI1, I2,

e. I5g are members of *L*2. Therefore, keep fI1, I2, I5g in *C*3. The 2-item subsets of fI1, I3, I5g are fI1, I3g, fI1, I5g, and fI3, I5g. fI3, I5g is not a member of *L*2, and so it is not frequent. Therefore, remove fI1, I3, I5g from*C*3. The 2-item subsets of fI2, I3, I4g are fI2, I3g, fI2, I4g, and fI3, I4g. fI3, I4g is not a member of *L*2, and so it is not frequent. Therefore, remove fI2, I3, I4g from*C*3.

f. The 2-item subsets of fI2, I3, I5g are fI2, I3g, fI2, I5g, and fI3, I5g. fI3, I5g is not a member of *L*2, and so it is not frequent. Therefore, remove fI2, I3, I5g from*C*3.

g. The 2-item subsets of fI2, I4, I5g are fI2, I4g, fI2, I5g, and fI4, I5g. fI4, I5g is not a member of *L*2, and so it is not frequent. Therefore, remove fI2, I4, I5g from*C*3.

h. Therefore, *C*3 = ffI1, I2, I3g, fI1, I2, I5gg after pruning.

8. The algorithm uses *L*3 on *L*3 to generate a candidate set of 4-itemsets, *C*4. Although the join results in ffI1, I2, I3, I5gg, this itemset is pruned because its subset ffI2, I3, I5gg is not frequent. Thus, *C*4 = f, and the algorithm terminates, having found all of the frequent itemsets.

Step 1 of Apriori finds the frequent 1-itemsets, *L*1.

In steps 2 to 10, $Lk\Box 1$ is used to generate candidates *Ck* in order to find *Lk* for *k* _ 2. The apriori gen procedure generates the candidates and then uses the Apriori property to eliminate those having a subset that is not frequent (step 3). This procedure is described below. Once all of the candidates have been generated, the database is scanned (step 4). For each transaction, a subset function is used to find all subsets of the transaction that are candidates (step 5), and the count for each of these candidates is accumulated (steps 6 and 7). Finally, all of those candidates satisfying minimum support (step 9) form the set of frequent itemsets, *L* (step 11). A procedure can then be called to generate association rules from the frequent itemsets. Such a procedure is described in Section 5.2.2.

The apriori gen procedure performs two kinds of actions, namely, join and prune, as described above. In the join component, $Lk\Box 1$ is joined with $Lk\Box 1$ to generate potential candidates (steps 1 to 4). The prune component (steps 5 to 7) employs the Apriori property to remove candidates that have a subset that is not frequent. The test for infrequent subsets is shown in procedure has infrequent subset.

We can't apply association rules; we got most frequent item sets. In our proposal system we can get all frequent items in order and also apply negative association rules.
In order to find frequent items we taken the given transactional data sets. Let us take the example in [1],[2], The transactional data for all electronics organization.

### III. ALGORITHM

//freaquecy count of words in a given text

```
Procedure bst()
    begin
    public:
    struct node
    {
    char info[19];
    int c;
    node *lp;
    node *rp; }*root;
        bst()
            begin
            root ← NULL;
            end
    Procedure insert()
        begin
        char ele[19];
        int x;
    node *new1=new node();
    write "enter word to insert"
        read(ele);
        new1->info ← ele;
```

```
            new1->c←1;
            new1->lp←NULL;
            new1->rp←NULL;
              if(isempty())
                begin
              root=new1;
                 end
                 else
                begin
           node *temp,*t1;
              temp←root;
             //travelling

         while(temp!=NULL)
                begin
              t1←temp;
    x=comparing two strings(ele,temp->info);
                if(x==0)
                 begin
         temp->c ← temp->c+1;
             temp ← NULL;
                  end
                  else
                 begin
                 if(x>0)
            temp←temp->rp;
                  else
            temp←temp->lp;
               else end
                  end
                 if(x<0)
            t1->lp←new1;
              else if(x>0)
            t1->rp←new1;
                  else
             delete new1;
              else end
          end procedure insert
           procedure inorder
                 begin

             node *temp;
             node *s[19];
             int top←-1;
             int flag←0;
            temp←root;
              while(!flag)
                 begin
          while(temp!=NULL)
                 begin
             top←top+1;
             s[top]←temp;
           temp←temp->lp;
                  end
               if(top!=-1)
```

```
                 begin
             temp←s[top];
             top←top-1;
           //the words are
           write(temp->info);
            write(temp->c);
           temp←temp->rp;
                  end
                  else
                 flag=1;
                else end
                  end

            procedure isempty
                 begin
           return(root==null)
             end procedure
```

IV.RESULTS:

//**The text is:**
//i reach my goal by my uncompromised practice
THE FREAQUENCY COUNT OF WORDS IN A GIVEN TEXT IS
The words are:      by
The word count=    1
The words are:    goal
The word count=    1
The words are:      i
The word count=    1
The words are:     my
The word count=    2
The words are:   practice
The word count=    1
The words are:    reach
The word count=    1
The words are:   uncompromised
The word count=    1
1.insert
2.frequent count inorder
3.exit
enter u r choice      3

V.CONCLUSION:

In order to make more efficient we need to use balanced binary search tree concepts such as AVL Trees, Splay trees, Red – Black trees. By using these concepts always in worst case time complexity also becomes to O(logn) for all operations. We expect hashing to outperform balanced search trees when the desired operations are search, insert and delete. Balanced search trees are recommended only in time critical applications in which we must guarantee that no dictionary operation ever takes more than a specified amount of time. Balanced search trees are also recommended when the search and delete operations are done by rank.

| Method | Worst Case | | | Expected | | |
|---|---|---|---|---|---|---|
| | Search | Insert | Delete | Search | Insert | Delete |
| Sorted Array | Log n | n | n | Log n | n | n |
| Sorted Chain | n | n | n | n | n | n |
| Skip Lists | n | n | n | Log n | Log n | Log n |
| Hash Table | n | n | n | 1 | 1 | 1 |
| Binary Search Tree | n | n | n | Log n | Log n | Log n |
| AVL Tree | Log n | Log n | Log n | Log n | Log n | Log n |
| Red-Black Tree | Log n | Log n | Log n | Log n | Log n | Log n |
| Splay Tree | n | n | n | Log n | Log n | Log n |
| B-Trees | Log n | Log n | Log n | Log n | Log n | Log n |